  \documentclass[12pt]{article} 

  \usepackage{tikz,natbib}
  \input{jmatharxiv.sty}
  \newcommand{\colk}{\color[named]{black}}
   \renewcommand{\colb}{\colk}
   
   \renewcommand{\colg}{\colk}
 
\begin{document}
\setlength{\baselineskip}{14pt}    

\par
\par
\par
\title{\sffamily\colb
  Comment on arXiv:2202.01553:
  The Distribution of a Gaussian Covariate Statistic }
 \par
\par
\author{\sffamily\colb  Joe Whittaker, Lancaster University
  \footnote{joe.whittaker@lancaster.ac.uk}}
\maketitle{}
\paragraph{\sffamily\colb  Abstract}
  In regression Gaussian covariate p-values \citep{davies2022covariate} are used
  to control greedy forward subset selection by accounting for
  choosing the best when fitting many variables.
  Here we outline a simple proof of their Theorems 1 and 2, making slight
  alterations to simplify the exposition
  by including a new variable rather than
  excluding an included variable and some slight changes in notation.
\par
\subsection*{\sffamily\colb  Background }
The problem of variable selection in linear regression, especially in
high dimensions, is still the object of current interest.
Many sequential techniques depend on comparing sums of squares from
different fits.
Under the usual modelling assumption that each observation of the dependent
variable comes from a Normal distrbution with constant variance
the standard F-statistic for adding a single variable has an ${\colb \calF}$
distribution, with certain degrees of freedom, and a consequent p-value.
Adding a second variable also leads to an ${\colb \calF}$ distribution, with
different degrees of freedom, giving its p-value.
However, there is no easy theory for computing the p-value when the
first variable is chosen to be the better of the two, the one with the
smaller residual sum of squares.
The contribution of \citet{davies2022covariate} is to show that within
the framework of independent Gaussian covariate regression such
p-values can be calculated.
Importantly these p-values are valid whatever the sampling
distribution of the dependent and explanatory variables.
\par
\par
\subsection*{\sffamily\colb  Standard linear regression preliminaries}
The vector ${\colb \bmy}$ is a sample of size ${\colb n}$ observations 
with ${\colb k}$ covariates held in a matrix ${\colb X}$ of dimension
${\colb n\times{}k}$.
The least squares predictor of ${\colb \bmy}$ from a linear combination of the
columns of ${\colb X}$ is ${\colb P\bmy}$ where ${\colb P= X(X^{T}X)X^{T}}$ is the projection
matrix onto the space spanned by the observed ${\colb X}$ variables and ${\colb I-P}$
is the residual space orthogonal to this.
The residuals are
{\colb\begin{eqnarray}\colb\label{eq:r}
\bmr &=&(I-P)\bmy
\end{eqnarray}}\blank\newline
with residual sum of squares  ${\colb \bmr^{T}\bmr= \bmy^{T}(I-P)\bmy}$.
The standard analysis of variance decomposition of the total sum of
squares is 
${\colb \bmy^{T}\bmy = \bmy^{T}P\bmy  +\bmy^{T}(I-P)\bmy}$,
and the F statistic for assessing the contribution of all covariates
is based on the ratio of these two terms.
\par
Consider whether an additional variable ${\colb \bmz}$ should be included
in the regression when the ${\colb X}$ variables are already included.
The regression of ${\colb \bmy}$ on ${\colb X}$ and ${\colb \bmz}$ can be acheived by first
fitting ${\colb X}$ and then regressing the residuals of ${\colb \bmy}$ from ${\colb X}$ on
the residuals of ${\colb \bmz}$ from ${\colb X}$.
Put
{\colb\begin{eqnarray}\colb\label{eq:s}
\bms &=&(I-P)\bmz
\end{eqnarray}}\blank\newline
then the projection onto the space spanned by the additional variable is
${\colb J(\bms)=\bms\bms^{T}/\bms^{T}\bms}$
and the sum of squares  decomposition of the ${\colb \bmy}$ residuals is
{\colb\begin{eqnarray}\colb\label{eq:decompr}
\bmr^{T}\bmr  &=& \bmr^{T}J(\bms) \bmr + \bmr^{T}(I-J(\bms))\bmr.
\end{eqnarray}}\blank\newline
The ratio of the two residual sums of squares
to assess  whether ${\colb \bmz}$ contributes to
the regression is
{\colb\begin{eqnarray}\colb\label{eq:Brs}
  B(\bmr,\bms)  &=&     \bmr^{T}(I-J(\bms))\bmr/\bmr^{T}\bmr.
\end{eqnarray}}\blank\newline
\par {We outline the derivation of the distribution of ${\colb B}$ under the
  standard model.}
One (of several) formulations of the standard model is the 
assumption that ${\colb \bmY\sim\calN(X\beta,\sigma^2I)}$ where the
capitalisation makes it clear that ${\colb \bmY}$ is random and ${\colb \bmy}$ is
a particular realisation.
Furthermore this  distribution remains the same
under the supposition that 
 the additional variable ${\colb \bmz}$ has no effect on the distribution of ${\colb Y}$.
From (\ref{eq:r}) 
${\colb \bmR = (I-P)\bmY}$
so that
{\colb\begin{eqnarray}\colb\label{eq:R}
\bmR\sim\calN(\bmzero,\sigma^2(I-P)).
\end{eqnarray}}\blank\newline
Returning to the quadratic forms in the decomposition at
(\ref{eq:decompr}) note that ${\colb J}$ and
${\colb I-J}$ are projections and are orthogonal, so by Cochran's theorem, they
have independent ${\colb \chi_{}^2}$ distributions with the appropriate
degrees of freedom, whatever ${\colb \bmz}$.
Consequently the ratio of the two residual sums of squares
{\colb\begin{eqnarray}\colb\label{eq:BRs}
 B(\bmR,\bms)\sim\calB_{((n-k-1)/2,1/2)},
\end{eqnarray}}\blank\newline
the Beta distribution,
under the assumption that ${\colb \bmY}$ does not depend on ${\colb \bmz}$. 
\par
\section*{\sffamily\large\colb The distribution of ${\colb B}$ in the Gaussian covariate framework }
The alternative Gaussian covariate framework supposes that given ${\colb X}$
and ${\colb \bmy}$, which may be fixed or random, the observations ${\colb \bmz}$ are
sampled from the Normal distribution,
${\colb \bmZ\sim\calN(\bmzero,\sigma_z^2I)}$, and so are independent of both
${\colb X}$ and ${\colb \bmy}$,
In this distribution-free framework \cite{davies2022covariate}
showed that, in parallel  to    (\ref{eq:BRs}),
{\colb\begin{eqnarray}\colb\label{eq:BrS}
  B(\bmr,\bmS)      \sim\calB_{((n-k-1)/2,1/2)},
\end{eqnarray}}\blank\newline
as well.
Here is a simple proof.
\par
First note that   from  (\ref{eq:Brs}) the statistic ${\colb B}$ satisfies
{\colb\begin{eqnarray}\colb 
  B(\bmr,\bms)  &=& B(\bms,\bmr), \jand   \label{eq:sym}  \\
  B(a\bmr,b\bms)  &=& B(a\bms,b\bmr)  \label{eq:scale}
\end{eqnarray}}\blank\newline 
where ${\colb a,b}$ are positive scalars.
So ${\colb B}$ is symmetric in ${\colb \bmr}$ and ${\colb \bms}$, and is scale invariant.
Symmetry is because
  ${\colb B(\bmr,\bms)
  = \bmr^{T}(I-J(\bms))\bmr/ \bmr^{T}\bmr     = 1 - (\bmr^{T}\bms)^2/(\bms^{T}\bms {\,} \bmr^{T}\bmr)}$,
which uses   ${\colb \bmr^{T}J(\bms)\bmr = (\bmr^{T}\bms)^{2}/{\bms^{T}\bms }}$
and is symmetric.
The right hand side is a function of the partial
correlation between ${\colb \bmy}$ and ${\colb \bmz}$ having adjusted for ${\colb X}$,
and is scale invariant.
\par
The proof continues
{\colb\begin{eqnarray*} 
  B(\bmr,\bmS)
  &=& B(\bmS,\bmr), \mbox{ symmetry,}  \\ 
  &=& B(\sigma_{\colg z}^{-1}\bmS,\bmr),     \mbox{ invariance,} \\ 
  &\stackrel{d}{=}& B(\sigma^{-1}\bmR,\bmr),
    \mbox{ distributional equivalence,} \\ 
  &=& B(\bmR,\bmr),\mbox{ invariance.}
\end{eqnarray*}}\blank 
The distributional equivalence is because under the standard model
${\colb \bmR}$ has the Normal distrbution (\ref{eq:R}) and from (\ref{eq:s})
within the independent Gaussian covariate framework  
${\colb \bmS\sim\calN(\bmzero,\sigma_Z^2(I-P))}$.
\par
\section*{\sffamily\large\colb  Corollary: application to  p-values }
The relationship between the Beta and the F distributions give 
an alternative way to express (\ref{eq:BRs}): the F statistic 
  ${\colb F(\bmr,\bms) =  {(n-k-1)(1-B(\bmr,\bms))}/{B(\bmr,\bms)}}$
 is distributed under the standard model as
{\colb\begin{eqnarray}\colb\label{eq:FRs}
  F(\bmR,\bms)  &\sim&\calF_{\colg (1,n-k-1)}.
\end{eqnarray}}\blank\newline
\par
Let the subscript `obs' denote  a specific numerical value for the
subscripted variable.
Hence the standard p-value, for including  ${\colb \bmz}$ in the regression
when ${\colb \bmz_{obs}}$ is observed, is
{\colb\begin{eqnarray}\colb 
  p_{\colg R}(F_{\colg obs})
    &=&  Pr[F(\bmR,\bms_{obs})>F_{\colg obs}]\nonumber\\ 
    &=&   1-\calF{}_{\colg (1,n-k-1)}(F_{\colg obs})  \label{eq:pvalR}
\end{eqnarray}}\blank\newline
using (\ref{eq:FRs}).
Computing ${\colb  p_{\colg R}(F_{\colg obs})}$ requires values for ${\colb \bmy_{obs}}$,
${\colb X_{obs}}$, ${\colb \bmz_{obs}}$ giving ${\colb \bmr_{\colg obs}}$, ${\colb \bms_{obs}}$ to get
${\colb F_{\colg obs}}$ and allowing the evaluation of  (\ref{eq:pvalR}).
\par
In the Gaussian covariate framework of the previous
section we have shown  that
{\colb\begin{eqnarray*} 
p_{\colg S}(F_{\colg obs}) =
Pr[F(\bmr_{\colg obs},\bmS)>F_{\colg obs}]  =   1-\calF{}_{\colg (1,n-k-1)}(F_{\colg obs})
\end{eqnarray*}}\blank
as well, so that finally
{\colb\begin{eqnarray}\colb\label{eq:pvalS}
  p_{\colg S}(F_{\colg obs}) =  p_{\colg R}(F_{\colg obs}),
\end{eqnarray}}\blank\newline
the p-values are the same.
\par
Now consider observing two additional variables, ${\colb \bmz_{1}}$ and ${\colb \bmz_{2}}$,
and evaluating the p-value for selecting the better of the two. 
 The Gaussian covariate framework  supposes that these
covariates are observations from Normal distributions, identical to
that associated with ${\colb z}$ above, and also are mutually independently.
Put ${\colb F^{12}_{\colg obs}=\max(F^{1}_{\colg obs},F^{2}_{\colg obs})}$ where the superscript indexes the variate, then the p-value,   ${\colb p^{12}_{\colg S}(F^{12}_{\colg obs})}$, is
{\colb\begin{eqnarray*} 
Pr[\max(F(\bmr_{\colg obs},\bmS_{1}),F(\bmr_{\colg obs},\bmS_{2})) > F^{12}_{\colg obs}]
  &=&
1-Pr[
  F(\bmr_{\colg obs},\bmS_{1})<F^{12}_{\colg obs} \mbox{ and }
  F(\bmr_{\colg obs},\bmS_{2})<F^{12}_{\colg obs}
]
\\
 &=& 1- (  1-\calF{}_{\colg (1,n-k-1)}(F^{12}_{\colg obs}))^2  \\
 &=& 1- (1- p_{\colg S}(F^{12}_{\colg obs})  ))^2.
\end{eqnarray*}}\blank
 \par
On the other hand  the computation of 
${\colb Pr[
  F(\bmR_{1},\bms_{1,obs})<F^{12}_{\colg max} \mbox{ and }
  F(\bmR_{2},\bms_{2,obs})<F^{12}_{\colg max}
]}$
where ${\colb \bmR_{1}}$ and ${\colb \bmR_{2}}$ are the  residuals of
${\colb Y}$ from the  respective covariates
is very much more difficult.
\par
The calculation generalises immediately to considering ${\colb q}$ extra covariates
rather than 2.
\paragraph{\sffamily\colb Acknowledgements:} Thanks go to Laurie Davies for encouragement to put these thoughts to writing and to Technical Support at the
School of Mathematical Sciences, Lancaster University,
\bibliography{joe2025proof}

\end{document}